\documentclass[12pt]{article}
\begin{document}
\font\bss=cmr12 scaled\magstep 0
\title{\vspace{2mm} On integration of the closed KdV dressing chain}

\author{\bss Yu. Brezhnev\dag, S. B. Leble\ddag , \\
\small
\ddag Faculty of Applied Physics and Mathematics\\
\small Technical University of Gdan'sk,  \\
\small  ul. G.Narutowicza, 11/12 80-952, Gdan'sk-Wrzeszcz, Poland,\\
\small   email leble@mifgate.pg.gda.pl\\
\small and\\
\small \dag \ddag Kaliningrad State University, Theoretical Physics Department,\\
\small Al.Nevsky st.,14, 236041 Kaliningrad, Russia.\\} \vskip
1.0true cm

\date {}
\renewcommand{\abstractname}{\small Abstract}
\maketitle
\begin{abstract} The  dressing chain equations  for the second
order Sturm-Liouville differential operators is integrated. The
simplest closure at the third step is linked to the spectral curve
of the genus 1 and the explicit solution in elliptic Weierstrass
functions  is given. Dependence on integrals that enter the
bihamiltonian structure is specified.

\end{abstract}

\thispagestyle{empty}

\section{Introduction}

This introduction is a citation from the paper \cite{VSha}. The
Darboux transformation scheme of dressing  you can found at
 \cite{L, Mat, LeZa}, the chains were introduced at
\cite{We}, studied at \cite{Sha,{Novik}}. Consider a dressing
chain for a Sturm-Liouville differential operator
\begin{equation}
(\sigma_{n}+\sigma_{n+1})_{x}=\sigma_{n}^{2}-\sigma_{n+1}^{2}+\alpha
_{n} \label{1}
\end{equation}
It is connected to the the linear Schr\"{o}dinger equation
\begin{equation}
\Psi _{xx}=(q-\lambda )\Psi ,  \label{2}
\end{equation}
by $q_{n}=\sigma_{n_{x}}+\sigma_{n}^{2}+\mu _{n},\,\,\,\alpha
_{n}=\mu _{n}-\mu _{n+1}.$ Let us close the chain (\ref{1}) by the
condition
\begin{equation}
\sigma_{i}\equiv \sigma_{i+N},\,\,\mu _{i}\equiv \mu _{i+N},
\label{3}
\end{equation}
and assume that $\sum_{i=1}^{N}\alpha _{i}=0,$ then we have finite
dimensional dynamical system
\begin{equation}
(\sigma_{i}+\sigma_{i+1})_{x}=\sigma_{i}^{2}-\sigma_{i+1}^{2}+\mu
_{i}-\mu _{i+1},\,\,\,i=1,...,N.  \label{4}
\end{equation}
As was shown by Veselov and Shabat \cite{VSha}, for $N=2n+1$, it
is bi-Hamiltonian system of the following form in
$g_{i}=f_{i}+f_{i+1}$ coordinates:
\begin{equation}
\pi _{\lambda }dh_{\lambda }=0\Leftrightarrow
\begin{array}{l}
\pi _{0}dh_{0}=0 \\
\pi _{0}dh_{1}=K_{1}=\pi _{1}dh_{0} \\
\pi _{0}dh_{2}=K_{2}=\pi _{1}dh_{1} \\
\,\,\,\,\,\,\,\,\,\,\,\,\,\,\,\,\,\,\,\,\,\,\,\,\,\,\,\,\,\,\,\,\vdots  \\
\pi _{0}dh_{n}=K_{n}=\pi _{1}dh_{n-1} \\
\qquad \,\,\,\,\,\,\,\,\,\,\,\,\,\,\,\,\,0=\pi _{1}dh_{n},\,\,
\end{array}
\label{5}
\end{equation}
\begin{equation}
\{g_{i},g_{i-1}\}_{\pi _{0}}=1,  \label{6}
\end{equation}
\begin{eqnarray}
\{g_{i},g_{j}\}_{\pi _{1}} &=&(-1)^{j-i}g_{i}g_{j},\,\,\,\,\,\,\,\,\,\,j\neq
i\pm 1,  \nonumber \\
\{g_{i},g_{i-1}\}_{\pi _{1}} &=&g_{i}g_{i-1}+\beta _{i}  \label{7}
\end{eqnarray}
and the generating function for $h_{i}$ is
\begin{eqnarray}
\tau _{N} &=&\left[ \prod_{j=1}^{N}\left( 1+\zeta _{j+1}\frac{\partial ^{2}}{%
\partial g_{j}\partial g_{j+1}}\right) \right] \prod_{k=1}^{N}g_{k}
\nonumber \\
&=&(-1)^{N}(h_{0}\lambda ^{n}+h_{1}\lambda
^{n-1}+...+h_{n}=(-1)^{n}h_{\lambda },,\,\,\,\,\,\,\,\,\,\,\,\zeta
_{i}=\beta _{i}-\lambda .  \label{8}
\end{eqnarray}
This is the end of the citation with the only comment that the
generating function could be obtained as irreducible
representation of the symmetry group of the dressing chain
equations (\ref{1}) \cite{Lt}. More general symmetry is studied in
\cite{FSV}.
\section{Explicit formulas for the solutions of the the chain equations $%
(N=3)$}

If the system (\ref{1}) is interpreted as a result of
factorization of the operator SL of the equation (\ref{2}) and the
relation (\ref{3}) supplies the necessary condition for this. If,
further the factorization is linked to the DT, then $\mu$ is the
spectral parameter corresponding to the auxiliary spectral
function that parameterized the transformation. The parameter
$\alpha$ is zero in this case.

Let us consider equations
\begin{equation}
\begin{array}{rcl}
\sigma_1^{\prime} & = & \sigma_3^2-\sigma_2^2+\mu_3-\mu_2, \\
\sigma_2^{\prime} & = & \sigma_1^2-\sigma_3^2+\mu_1-\mu_3, \\
\sigma_3^{\prime} & = & \sigma_2^2-\sigma_1^2+\mu_2-\mu_1,
\end{array}
\end{equation}
that is equivalent to the system (\ref{4}) under the condition
(\ref{3}) with N=3. We use two integrals
\begin{equation}
\begin{array}{ccl}
C & = & \sigma_1+\sigma_2+\sigma_3 \\
A & = & g_1g_2g_3 +\mu_2 g_3+\mu_1 g_2+\mu_3 g_1= \\
&  & (\sigma_1+\sigma_2) (\sigma_2+\sigma_3)(\sigma_3+\sigma_1)+ \\
&  & \mu_2(\sigma_3+\sigma_1)+\mu_1(\sigma_2+\sigma_3)+\mu_3(\sigma_1+%
\sigma_2)
\end{array}
\end{equation}
From (1) one obtains
\[
{\sigma^{\prime}_1}^2=(\sigma_3^2-\sigma_2^2+\mu_3-\mu_2)^2
\]
Using the first equation (2) $\sigma_3=C-\sigma_1-\sigma_2$ we will obtain
the equation containing $\sigma^{\prime}_1, \sigma_2$. As the next step we
eliminate the remaining variable $\sigma_2$ with the help the second
integral in (2) and the first also. Thus, we get the equation
\begin{equation}
\begin{array}{ccl}
{\sigma^{\prime}_1}^2 & = & \sigma_1^4-2(C^2+\mu_3+\mu_2-2\mu_1)\sigma_1^2
\\
&  & -4(2\mu C-A) \sigma_1 +C^4+2(\mu_2+\mu_3+2\mu_1)C^2-4AC+(\mu_3-\mu_2)^2
\end{array}
\end{equation}
The last equation has the form
\begin{equation}\label{w}
  \left(\frac{\sigma_1}{dx}%
\right)^2 = \sigma_1^4-6a\sigma_1^2+4b\sigma_1+d,
\end{equation}
 where $a,b,d$ are constants defined by the previous
expression (3). The extra multipliers 6,4 have been included for a
convenience. The relation (\ref{w}) is an elliptic curve in
variables $(\sigma^{\prime}_1, \sigma_1)$ and therefore it is
uniformised by elliptic functions. Let us built the invariants
(capital letters are chosen that to distinguish them from
variables $g_j$ of the chain)
\[
G_2=d+a^2\qquad G_3=a^3-b^2-ad
\]
So, the pair $(b,a)$ is a point on a curve
\[
b^2=4a^3-G_2a-G_3.
\]
Therefore, there exists a parameter $\nu$ such that the following
equations will be hold:
\[
b=\wp^{\prime}(2\nu),  \qquad a=\wp(2\nu).
\]
This means that we take three new parameters $G_2,G_3,\nu$ instead of old
ones $a,b,d$ depended on five parameters of the chain: $(\mu_1,\mu_2,%
\mu_3,A,C)$. Now we may write
\begin{equation}\label{main}
  \left(\frac{d\sigma_1}{dx}%
\right)^2 = \sigma_1^4-6\wp(2\nu)\sigma_1^2+
4\wp'(2\nu)\sigma_1+(G_2-3\wp(2\nu)),
\end{equation}
that yields
\[
\sigma_1(x)=\zeta(x+\nu+x_0;G_2,G_3)
-\zeta(x-\nu+x_0;G_2,G_3)-\zeta(2\nu;G_2,G_3).
\]
Note, the $\sigma_1$ is not Weierstrass's $\sigma$-function in the
theory of elliptic functions, but $\zeta$ is standard Weierstrass
$\zeta$-function. The solution $\sigma_1(x)$ contains three
arbitrary constants (according to the third order of equations
(1)) $G_2,G_3,x_0$ which are, in its turn, defined by five ones
$\mu_j,A,C$ in explicit  but transcendental way. Parameter $\nu$
is not exceptional
\[
\nu=\frac12 \wp^{-1}\left(\frac{\mu_3+\mu_2-2\mu_1+C^2}{3}\right),
\]
where $\wp^{-1}$ denotes an elliptic integral of the first kind (inversion
of elliptic function $\wp$).

\textbf{Remark 1:} Just obtained solution solution is exactly
logarithmic derivative of the $\Psi$-function for the 1-gap Lame
potential
\[
\Psi^{\prime\prime}-2\wp(x)\Psi=\lambda \Psi\qquad \Psi=\frac{%
\sigma(\alpha-x)}{\sigma(\alpha)\sigma(x)} e^{\zeta(\alpha) x}, \qquad
\lambda=\wp(\alpha).
\]
and distinguished from the solution
\[
\frac{\Psi^{\prime}}{\Psi} =\zeta(\alpha-x) -\zeta(\alpha)- \zeta(x)
\]
by the shift of the spectral parameter $\alpha$.

\textbf{Remark 2:} If one interests in $\sigma_i$ (\ref{w}) in a
connection with KdV equation theory, the dependence on time could
be obtained using the t-chains \cite{Lt}, obtained by means of
MKdV equation for $\sigma$ and conservation laws \cite{Wad}.

The work is supported by the Polish Ministry of Scientific
Research and information Technology grant PBZ-Min-008/P03/2003.


\begin{thebibliography}{99}
\bibitem{VSha}  Veselov A, Shabat A, 1993, Funk. analiz i pril.\textbf{27}
p.1
\bibitem{L}  Leble S, 1991 Darboux Transforms Algebras in 2+1 dimensions in
\textit{Proc. of 7th Workshop on Nonlinear Evolution Equations and Dynamical
Systems ed. M Boiti et al, World Scientific, Singapore\/} p.53-61. \textit{%
Computers Math. Applic.}, \textbf{35} pp. 73-81, (1998).

\bibitem{Mat}  Matveev V B 1998 Darboux Transformations in Associative Rings
and Functional-Diffrence Equations ed J Harnad and A Kasman "The Bispectral
Problem" AMS series CRM PROCEEDINGS AND LECTURE NOTES v.14, p.211-226.

\bibitem{LeZa}  Zaitsev A, Leble S, 1999 Preprint 12.01.1999
math-ph/9903005; 2000 Reports on Math. Phys. \textbf{46} 165-167.


\bibitem{We}  J. Weiss, 1986, J. Math. Phys,\textbf{27} p.2647.

\bibitem{Sha}  Shabat A 1992 The infinite-dimensional dressing dynamical
systems, Inverse Problems, \textbf{8} 303-308

\bibitem{Novik}  Novokshenov V Yu 1995 Reflectionless potentials and soliton
series of the nonlinear Schr\"odinger equation, Physica D \textbf{87}
109-114.

\bibitem{Bat}  Bateman H Erdeli A 1955 Higher Transcendental Functions v3,
Mc Graw-hill inc.

\bibitem{FSV}  Fordy A Shabat A Veselov A 1995 Factorization and Poisson
correspondence Teor Mat Fiz \textbf{105} 225-245.

\bibitem{Wad}  Wadati M 1975 J of Phys. Soc. Japan  673-680.

\bibitem{Lt} S. Leble Darboux-covariant differential-difference operators and
dressing chains. ArXive.
\end{thebibliography}
\end{document}